\documentclass{ws-procs975x65}

\usepackage{multirow}

\usepackage[utf8]{inputenc}

\usepackage{xcolor}
\xdefinecolor{mylinkcolor}{rgb}{0,0,0.5}
\usepackage[
	bookmarksnumbered, bookmarksopen, bookmarksopenlevel=2,
	breaklinks=true,
	colorlinks=true, filecolor=mylinkcolor, citecolor=mylinkcolor,
	linkcolor=mylinkcolor, urlcolor=mylinkcolor, menucolor=mylinkcolor,
]{hyperref}

\begin{document}

\title{Spin effects on the dynamics of compact binaries}

\author{Jan Steinhoff}

\address{Max Planck Institute for Gravitational Physics (Albert Einstein Institute)\\
Am M{\"u}hlenberg 1, D-14476 Potsdam-Golm, Germany\\
E-mail: jan.steinhoff@aei.mpg.de}

\begin{abstract}
Compact binaries are the most promising source for the advanced gravitational wave detectors,
which will start operating this year. The influence of spin on the binary evolution is an
important consequence of general relativity and can be large. It is argued that the spin
supplementary condition, which is related to the observer dependence of the center,
gives rise to a gauge symmetry in the action principle of spinning point-particles. These
spinning point-particles serve as an analytic model for extended bodies. The internal
structure can be modelled by augmenting the point-particle with higher-order multipole
moments. Consequences of the recently discovered universal (equation of state independent)
relations between the multipole moments of neutron stars are discussed.
\end{abstract}


\bodymatter

\section{Conservative Spin Effects to Fourth Post-Newtonian Order}
An important source for the advanced ground-based gravitational wave detectors\cite{LIGO,VIRGO,KAGRA}
are inspiraling and merging binaries of compact stellar-mass objects. An analytic description
of the inspiral phase and the emitted gravitational waves is given by the post-Newtonian (PN) approximation,
which is a weak field and slow motion approximation. The description of the conservative part of
the motion was completed to 4PN order for (nonrotating) point masses recently.\cite{Damour:2014jta, Bernard:2015njp}

However, it is important to include the effects of the angular momenta of the bodies, i.e., theirs spins,
to the same order of approximation.
% In particular when one or both of the objects are black holes,
% one should include arbitrary spin magnitudes and orientations in the GW models.
Also this was achieved recently.
All the work that went into this can be summarized in the following table, sorted by PN and spin order S:
\begin{center}
\begin{tabular}{rcccccc}
           & 1.5PN & 2PN & 2.5PN & 3PN & 3.5PN & 4PN \\ \toprule
\multirow{2}{*}{S} & H\cite{Tulczyjew:1959,Barker:1975ae,Barker:OConnell:1979,Damour:1982,Damour:1988mr} && H\cite{Damour:2007nc,Steinhoff:2008zr,Perrodin:2010dy,Porto:2010tr,Levi:2010zu} && H\cite{Hartung:2011te, Hartung:2013dza, Levi:2015uxa} & \\
	   & E\cite{D'Eath:1975vw,Thorne:1984mz} && E\cite{Tagoshi:2000zg, Faye:2006gx} && E\cite{Marsat:2012fn, Bohe:2012mr} & \\ [0.7ex]
\multirow{2}{*}{S$_1^2$} && H\cite{Barker:1975ae,Barker:OConnell:1979,Poisson:1997ha} && H\cite{Porto:2008jj, Steinhoff:2008ji, Hergt:2010pa} && H\cite{Levi:2015ixa,Levi:2015msa} \\
	   && E\cite{D'Eath:1975vw,Thorne:1984mz} && E\cite{Bohe:2015ana} && \\ [0.7ex]
\multirow{2}{*}{S$_1$S$_2$} && H\cite{Barker:1975ae,Barker:OConnell:1979} && H\cite{Steinhoff:2007mb, Porto:2008tb, Levi:2008nh} && H\cite{Hartung:2011ea,Levi:2011eq,Hartung:2013dza} \\
          && E\cite{D'Eath:1975vw,Thorne:1984mz} && P\cite{Porto:2006bt}E\cite{Bohe:2015ana} && \\ [0.7ex]
\multirow{2}{*}{S$^3$}   &&&&& H\cite{Hergt:2007ha, Hergt:2008jn, Levi:2014gsa, Marsat:2014xea} & \\
        &&&&& E\cite{Marsat:2014xea}P\cite{Vaidya:2014kza} & \\ [0.7ex]
\multirow{2}{*}{S$^4$}   &&&&&& H\cite{Levi:2014gsa} \\
        &&&&&& P\cite{Hergt:2007ha,Hergt:2008jn,Vaidya:2014kza} \\
$\vdots$   &&&&&& $\ddots$
\end{tabular}
\end{center}
Here H stands for Hamiltonians or potentials, E refers to results for equations of motion, and P denotes partial/incomplete results. Some of the references at S$_1^2$ order are only valid for black holes but not for generic bodies like neutron stars.\cite{D'Eath:1975vw,Hergt:2007ha,Hergt:2008jn, Steinhoff:2008ji}

A variety of different formalisms was used, in particular for the more recent works.\cite{Blanchet:2013haa,Steinhoff:2010zz,Porto:2008tb,Levi:2015msa}
This is important, since the calculations bear
conceptual and technical difficulties, which makes independent checks mandatory.
For instance, a disagreement in the 4PN point-mass results\cite{Damour:2014jta, Bernard:2015njp} still
needs to be reconciled.
In the case of spinning bodies, some works were dedicated to establish connections between
the formalisms and the comparison of results.\cite{Hergt:2011ik,Levi:2014sba}
All results to 4PN with spin were checked by independent collaborations and using independent methods,
except for the next-to-next-to-leading order S$_1^2$ potential.\cite{Levi:2015ixa}
(The S$^4$ partial result\cite{Hergt:2007ha,Hergt:2008jn,Vaidya:2014kza} taken together form a
complete result in agreement with Ref.~\refcite{Levi:2014gsa}.)
It is also the only result where a Hamiltonian form is missing, from which, e.g., the gauge
invariant binding energy can be derived.

\section{Spin Gauge Symmetry in an Action Approach}
It is convenient to derive the above results from an action principle.
For a review of action principles with spin, see, e.g., Refs.~\refcite{Blanchet:2013haa,Steinhoff:2014kwa}.
A more recent concept is the introduction of a so called spin gauge symmetry
at the level of the action.\cite{Steinhoff:2015ksa, Levi:2015msa}
The motivation for this symmetry arises from the observer-dependence of the
center in relativity. This is best explained by the following figure.

\begin{center}
\begin{tabular}{p{3cm}p{9cm}}
\raisebox{-0.94\height}{\includegraphics{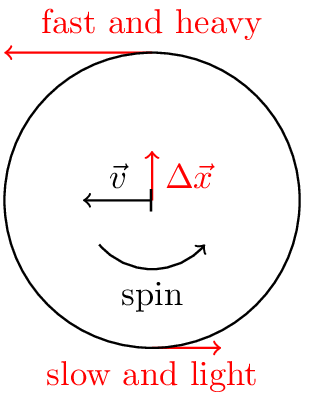}}
&
Figure:
Consider a spinning spherical symmetric object, which moves with a velocity $\vec{v}$ to the left.
Since the observer sees the upper hemisphere moving faster than the lower one,
the former acquires a larger relativistic mass than the latter.
The observed center of mass is therefore shifted by $\Delta \vec{x}$ away from the geometric center,
see, e.g., Ref.~\refcite{Fleming:1965}.
This shift depends on $\vec{v}$, that is, it is observer dependent. In the rest-frame, the
center of mass coincides with the geometric center.
\end{tabular}
\end{center}

This ambiguity in the center becomes problematic once a quadrupole or higher multipoles
of the body are taken into account, since the definition of the multipoles
hinges on the center as a reference point. One way to overcome this problem is to pick
the rest-frame center of mass, which is singled out since the rest frame provides an intrinsically
defined observer. But other centers are useful, too, for instance for Hamiltonian
descriptions. Now, any choice of a reference point in a body describes the same physical
situation, i.e., it can be understood as a gauge choice. It is then natural to expect that
this gauge freedom should correspond to a (gauge) symmetry in an action principle,
as implemented in Refs.~\cite{Steinhoff:2015ksa, Levi:2015msa}.
Interestingly, this action is supported on the rest-frame center worldline and the
shift is encoded through a time derivative of the linear momentum in the action.

The choice of center is usually encoded in a condition on the spin 4-tensor. This is
called the spin supplementary condition. A generalized version of this condition is
the generator of spin gauge transformations in the special relativistic case.\cite{Steinhoff:2015ksa}

\section{Finite Size, Multipole Moments, and Universal Relations}
The (effective) action principle for spinning bodies is a point-particle action with support on a worldline.
An interesting question is therefore how the finite size of the body is taken into account.
On large scales, the internal structure is encoded in the multipole moments of the bodies.
These are represented in the action through nonminimal coupling terms. Following an effective field
theory philosophy, one can construct all possible terms compatible with the symmetries, with
a constant for each term. The first terms in this expansion indeed correspond to the
quadruple, octupole, and hexadecapole\cite{Porto:2008jj,Levi:2015msa,Marsat:2014xea,Vaidya:2014kza}.

If all the constants in the action would be arbitrary, then one obtains
an undesirable enlargement of the parameter space for waveform models. Fortunately,
certain universal relations were discovered\cite{Yagi:2013bca,Yagi:2013awa} for slow rotating
compact objects (neutron stars, quark stars),
which are approximately independent of the nuclear equation of state. These
also hold for rapid rotation\cite{Pappas:2013naa,Chakrabarti:2013tca}
and for multipole moments up to the hexadecapole\cite{Pappas:2013naa,Yagi:2014bxa}.
This implies that the waveform model does approximately depend only on one additional
constant, which takes on different values for black holes and each neutron star model.
This even holds when tidal effects are taken into account.
Such a reduction of parameters is important, since the difficulties in extracting parameters
from gravitational waves grow significantly with the dimension of the parameter space.

%\section*{Acknowledgments}

%\bibliographystyle{utphys}

%\bibliography{references}
\providecommand{\href}[2]{#2}\begingroup\raggedright\endgroup

%\bibliography{referencesnotitle}
%\input{refsnotitle}

\end{document}